# Impurity-mediated early condensation of an atomic layer electronic crystal

Han Woong Yeom[1,2,*], Deok Mahn Oh[1,2], S. Wippermann[3], & W. G. Schmidt[4]

[1]*Centre for Artificial Low Dimensional Electronic Systems, Institute for Basic Science (IBS), 77 Cheongam-Ro, Pohang 790-784, Korea*

[2]*Department of Physics, Pohang University of Science and Technology, 77 Cheongam-Ro, Pohang 790-784, Korea*

[3]*Interface Chemistry and Surface Engineering Department, Max-Planck-Institute for Iron Research GmbH, Max-Planck-Straβe 1, 40237 Düsseldorf, Germany*

[4]*Lehrstuhl für Theoretische Physik, Universität Paderborn, 33098 Paderborn, Germany*

**While impurity has been known widely to affect phase transitions, the atomistic mechanisms have rarely been disclosed. We directly show in atomic scale how impurity atoms induces the condensation of a representative electronic phase, charge density wave (CDW), with scanning tunneling microscopy. Oxygen impurity atoms on the self-assembled metallic atomic wire array on a silicon crystal condense CDW locally even above the transition temperature. More interestingly, the CDW along the wires is induced not by a single atomic impurity but by the cooperation of multiple impurities. First principles calculations disclose the mechanism of the cooperation as the coherent superposition of the local lattice strain induced by impurities, stressing the coupled electronic and lattice degrees of freedom for CDW. This newly**

**discovered mechanism can widely be applied to various important electronic orders coupled to lattice, opening the possibility of the atomic scale strain engineering.**

Effects of impurity in phase transitions are important and widely studied. Impurity is generally believed to prohibit long range orders[1] or induce the condensation of low temperature phases below the transition temperature ($T_c$)[2,3]. However, the dominant majority of previous studies focused not on individual impurities but on statistical ensembles of them. Furthermore, while active impurities can even be single atoms or molecules, the direct atomic scale observation of the impurity-mediated condensation of phases has been rarely reported due both to the limited spatial resolution and dynamic nature of phase transition phenomena, leaving the microscopic mechanism elusive.

In the electronic transitions of condensed matter systems, the impurity effect is also believed to be crucial[4–6]. At low temperature the impurity and defect distributions were found correlated with the superconducting phase inhomogeneity, both for high and low temperature superconductors[4,5] and with domain structures of charge density wave (CDW) phases[6]. These phenomena would obviously be connected to the initial condensation of electronic phases with a finite density of atomic scale impurities which are usually ill characterized and whose microscopic interactions with electronic phases are not fully understood[4–6]. Moreover, the impurity-mediated condensation of these electronic phases has never been directly probed near $T_c$.

In this study, we made three important progresses. Firstly, we accessed near Tc to observe directly the condensation of CDW in atomic scale around well characterized and controllable impurities. Secondly, we found for the first time the atomic impurities enhancing the electronic order, inducing CDW above Tc. Finally, we made clear the microscopic mechanism of the early condensation through first principles calculations. The CDW system we choose is a quasi 1D metallic system formed on a semiconductor surface of Si(111)[7]. Indium atoms deposited at an elevated temperature self-organize into a perfectly ordered array of metallic atomic wires; two zigzag In chains form a unit wire separated by Si zigzag chains (Figs. 1b and 2a)[8]. These wires have three quasi 1D metallic bands, which, together with a strong coupling to the lattice, drive a metal-insulator transition into a periodically distorted insulating state at 125 K[7,9]. In this CDW phase, the dimerization along In zigzag chains and the sheer displacement between two neighboring chains result in the formation of the characteristic hexagon structure (Figs. 1c and 2b)[10].

This CDW system confined within a single atomic layer, an atomic layer electronic crystal, minimizes the interlayer coupling between layers of conventional bulk CDW materials. This makes the direct observation of the impurity effect substantially easier; the impurity can easily be introduced and exposed as adsorbates, which can be characterized fully and controlled systematically in stark contrast to impurities in bulk systems[11,12]. Indeed, various impurity atoms were investigated previously such as H, In, Pb, Na, and Co[13–17]. They commonly suppressed the local CDW order[15–17], causing the inhomogeneous phase separation and lowering $T_c$. This is consistent with the general wisdom of impurity effects[1–3] and what were observed in various electronic crystals[4,5]. However, they do not provide any new insight into the CDW condensation. In contrast, the CDW Tc was recently reported to increase proportional to the oxygen dose[16,17]. The pristine In wires were found to be close to the optimal doping and the electron or hole doping by oxygen adsorbates was found to be marginal[18]. This suggests the existence of a novel local impurity-CDW interaction beyond the doping effect[19].

Figure 1a shows the In wires at 140 K, above Tc of 125 K, with 3 Langmuir dose of $O_2$. The corrugation-less parts of the wires with the bright contrast are in the undistorted metallic phase as enlarged in Fig. 1b[20]. The short stripes and wider patches in the dark contrast are within the CDW state[20], as confirmed by the distinct STM images (Fig. 1c) of In hexagons. The CDW gap on these stripes and patches are confirmed as discussed below. Namely, CDW already condenses locally at 140 K above $T_c$. There is a linear relationship between the total area of the CDW condensates and the oxygen dose (Fig. 1d). This apparently indicates that the local CDW state is induced by oxygen impurities. The local CDW condensates exhibit only marginal fluctuations suggesting further that they are pinned by impurities.

The oxygen impurities are found as isolated dark or bright spots as shown in Figs. 1, 3a and 3b. The atomic structures of oxygen impurities were established in detail in previous works (Supplementary Fig. S1)[21,22]. The dark impurity is the dominant species, called $\alpha$, where the oxygen atom adsorbs on one In zigzag chain to bond with three neighboring In atoms (Figs. 2c and 3a). The minority species appearing as a bright spot results from an activated process of the oxygen incorporation below the In layer[21](Fig. 3b and supplementary Fig. S1). To our surprise, these isolated impurities and the isolated vacancy defects (Supplementary Fig. S2) do not induce CDW condensates but the ×2 lattice distortion appears with a decaying amplitude from them. These distortions are distinct from the CDW hexagon in topography (Figs. 3d, 3e, and 3f) and not insulating at all in spectroscopy (Fig. 3h). In stark contrast, in between two adjacent $\alpha$ impurities, the CDW hexagons are formed (Figs. 3d, 3e and 3f) with a clear energy gap (left panel of Fig. 3h), unlike the In atomic wire between two adjacent other oxygen impurities (Fig. 3g). We checked every isolated CDW stripes and found that the $\alpha$ impurity and another type of defects of the pristine surface (see Fig. 4c and supplementary Fig. S2)[23] couple with CDW. When two of them are within a distance of about 20 unit cells or shorter, they form CDW stripes (Supplementary Fig. S3).

Beyond single wires, small CDW patches can be formed as seeded by a CDW stripe and another $\alpha$ impurities in a neighboring wire as shown in Fig. 4a. This can be extended to wider CDW patches across a few wires with several $\alpha$ impurities clustered (Fig. 4b). For a larger patch, a CDW stripe (boxed in Fig. 4c) can be induced on a wire without an impurity when sandwiched between two CDW stripes bounded by impurities. These few cases can explain most of the impurity-CDW configurations (Supplementary Fig. S4). That is, there exist only two apparent interactions to condense CDW, the cooperation of two neighboring impurities within a single wire and the interwire coupling with neighboring CDW stripes. While the interwire interaction of the CDW in the quasi 1D systems is obviously expected and apparent in the 2D ordering below $T_c$[6], the microscopic mechanism underlying the cooperation of two impurities remains to be clarified.

In order to clarify this microscopic mechanism, we performed first principles calculations with single (Figs. 2a and 2b) and double (Figs. 2c, 2d, and 2e) impurities. As reported previously, isolated oxygen impurities induce the local $\times 2$ lattice distortions but not the CDW hexagons[21]. In clear contrast, two adjacent impurities can induce CDW hexagons depending on their distance. When two impurities have a distance of an even multiple of $a_0$ (Fig. 2d, $a_0$ is a Si lattice constant, 0.384 nm), being commensurate with the CDW periodicity, hexagons are readily formed but not in the other case of an odd multiple of $a_0$ (Fig. 2c), being incommensurate with CDW. In the former case, the lattice distortions of the neighboring oxygen impurities coherently overlap to make the hexagon formation energy lower. This is due to the fact that the CDW hexagon is based on the dimerization of In atoms of the outer In chains (In atoms bonded with Si zigzag chains)[9,10] and the oxygen adsorbate helps this dimerization by pulling two neighboring In atoms closer. The formation energy of hexagons is lowered by two cooperating oxygen atoms that are an even multiple of $a_0$ apart (8 and 30 meV, respectively for the configurations in Fig. 2d and 2e). In contrast, In hexagons are not even metastable when bracketed between O atoms that are an odd multiple of $a_0$ apart; the lattice distortions by two impurities quench each other (Fig. 2c). This is clearly observed in the experiment (Fig. 3c). Moreover, we also confirmed that the lattice distortion induced by impurities does not favor the hexagon formation, supporting consistently the experimental result (Fig. 3g and supplementary Fig. S5). Thus, the coupling mechanism of the oxygen impurity with the CDW is simply based on the proper bond length modification due to the impurity and the coherent superposition of the strain field imposed by multiple impurities. In the present case, a strain effect is very natural, since the CDW is based on the electron-lattice interaction.

To the best of our knowledge, there is no previous work to address the condensation of electronic crystals such as CDW and superconductivity[4,5] by impurities in atomic scale. A similar study was reported for a 2D surface phase transition on a Ge(111) surface[24], which was, however, shown to a simple order disorder transition in later studies[25]. For conventional superconductors, the magnetic impurity was observed to reduce Tc through the exchange scattering of cooper pairs[26,27]. Some impurities also suppress high-temperature superconductivity[26,27] but the microscopic mechanism is elusive as closely related to the origin of the superconductivity itself[5]. In these materials, the superconductivity is even entangled with the CDW order, which in turn closely coupled to impurities[28]. Therefore, the present case of the enhanced electronic order by atomic scale impurities is unique by itself and its transparent microscopic mechanism substantially widens our understanding of the interplay between impurities and electronic degrees of freedoms. The importance of the local strain filed was recently recognized also for two dimensional electronic systems of atomic crystals such as graphene[29] and surface states of topological insulators[30]. In those cases, however, the sources of local strains could not be controlled but just given as defects. In contrast, the present work with adsorbates provides controllability and, thus, steps forwards to the atomic scale strain engineering of such important electronic systems.

**Methods**

**Sample preparation.** The sample was prepared in the ultrahigh vacuum condition (UHV, a base pressure is $1.0 \times 10^{-10}$ torr). The clean Si(111)7×7 substrate was prepared by cycles of annealing and flashing up to 1,500 K of the n-doped Si(111) wafer. The array of In atomic wires were prepared by depositing one monolayer of In onto a Si(111)7×7 surface kept at 570 K[7]. Oxygen molecules were dosed on to the In atomic wire array at room temperature by backfilling the chamber through a precision leak valve. The $O_2$ dose is expressed in Langmuirs (1 L = $1.0 \times 10^{-6}$ torr s).

**STM measurement.** Scanning tunneling microscopy (STM) and spectroscopy (STS) measurements were performed using a commercial cryogenic UHV STM system (Unisoku) at 140 K and a base pressure $5.0 \times 10^{-11}$ torr. The electrochemically etched tungsten tips were cleaned by the e-beam heating in the UHV chamber. The STM topography was measured in a constant-current mode, and the STS, the d$I$/$dV$ signals, were obtained by a standard lock-in technique with a modulation of frequency $f$= 1 KHz and amplitude 30 mV$_{rms}$.

**DFT calculations.** Density functional theory calculations were performed using the local density approximation (LDA) as implemented in the Vienna *Ab-initio* Simulation Package (VASP). In the LDA calculations, indium 4d electrons were treated as core electrons. The surface was modeled using three Si bilayers within a 12×8 lateral unit cell. The Brillouin zone was performed using a k-point sampling equivalent to 256 points within a 1×1 unit cell.

**Acknowledgements** This work was supported by Institute for Basic Science (IBS-R014-D1). The numerical calculations were performed using grants of computer time from the Paderborn Center for Parallel Computing ($PC^2$) and the Hochstleistungs-Rechenzentrum Stuttgart (HLRS). The Deutsche Forschungsgemeinschaft (DFG) is acknowledged for financial support (FOR1700).

**Author contributions** H. W. Yeom established the basic concept, designed the experiments, and wrote the manuscript with other authors. D. M. Oh performed the STM experiments. S. Wippermann performed the theoretical calculations with the help of W. G. Schmidt.

**Competing Interests**  The authors declare that they have no competing financial interests.

**Correspondence** Correspondence and requests for materials should be addressed to Han Woong Yeom (email: yeom@postech.ac.kr).

**Figure 1 CDW of indium atomic wires with oxygen impurities. a**, The STM topographic image after 3 L oxygen dose onto the In atomic wires at 140 K (> $T_c$ = 125 K). The image shows the bright metallic parts (dashed box, enlarged in **b**) and dark local CDW condensates (box, enlarged in **c**) coexisting. The impurities are bright and dark spots (solid and dashed circles, respectively). Imaging conditions are $V_{bias}$ = -0.5 V and $I$ = 100 pA. **d**, The total area of the CDW domains as a function of the oxygen exposure.

**Figure 2 Atomic structures of In atomic wires with oxygen impurities.** The atomic structures of **a**, the metallic In wire above $T_c$ and **b**, the wire in the CDW state below $T_c$. The circles ovals represent the protrusions in filled state STM images shown in Fig. 1c The calculated structure of an In atomic wire with two $\alpha$ impurities odd multiple of $a_0$ apart ($5a_0$) which exhibits only the minor local distortion. The calculated structures with two impurities even multiple of $a_0$ apart ($4a_0$, **d** and **e**), which yield the hexagon CDW structure between impurities.

**Figure 3 Single and double oxygen impurities and the formation of the CDW condensates. a** and **b**, The STM images of oxygen impurities in two different adsorption structures, $\alpha$ and $\beta$, respectively, with lattice distortions around. **c**, **d**, **e**, and **f**, STM images of two adjacent oxygen impurities (indicated by arrows, $\alpha$ structure) without (**c**) and with ( **d**, **e**, and **f**) CDW hexagons (underlined in d and e) formed in between. The distance between the impurities are $13a_0$, $8a_0$, $14a_0$, and $16a_0$, respectively. **g**, In atomic wire

between two *β* impurities, which do not induce CDW (distance between two *β* impurities is 11$a_0$). Imaging conditions are $V_{bias}$ = -0.5 V and $I$ = 100 pA. **h**, Spatially and energetically resolved normalized differential conductance taken along the dashed arrows in **f** and **g**. The CDW energy gap (dashed lines) exists only for the CDW hexagons in between two *α* impurities.

**Figure 4 CDW patches formed by impurity clusters. a**, The small CDW patch composed two CDW stripes around three impurities. **b**, A larger patch with several impurities clustered and a single vacancy defect (yellow arrow). **c**, A much larger patch, which contain a CDW stripe (box) without any terminating impurity but sandwiched by neighboring CDW stripes. Imaging conditions are $V_{bias}$ = -0.5 V and $I$ = 100 pA

**Figures:**

**Fig. 1**

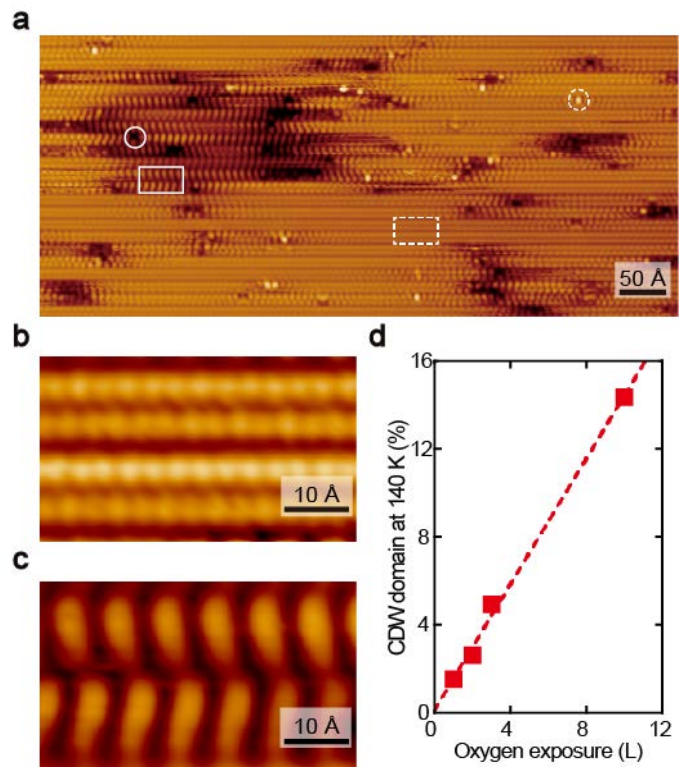

**Fig. 2**

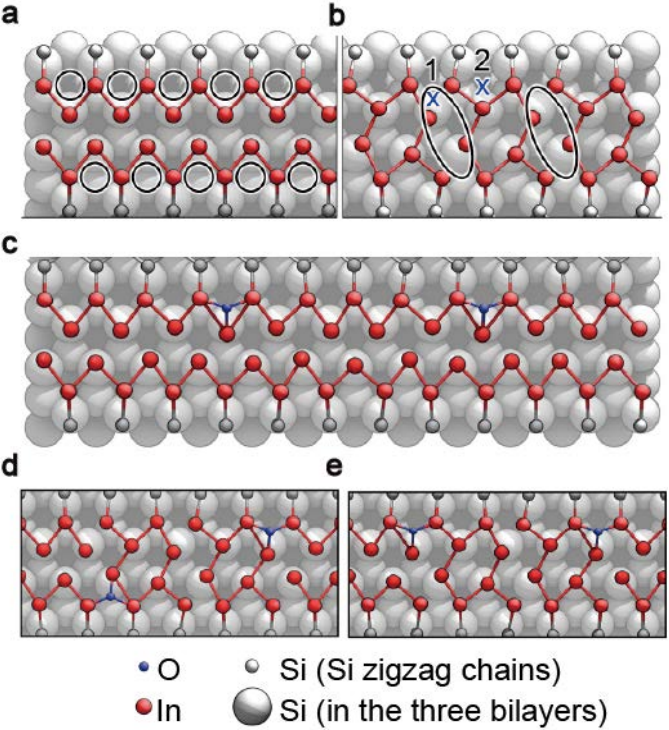

**Fig. 3**

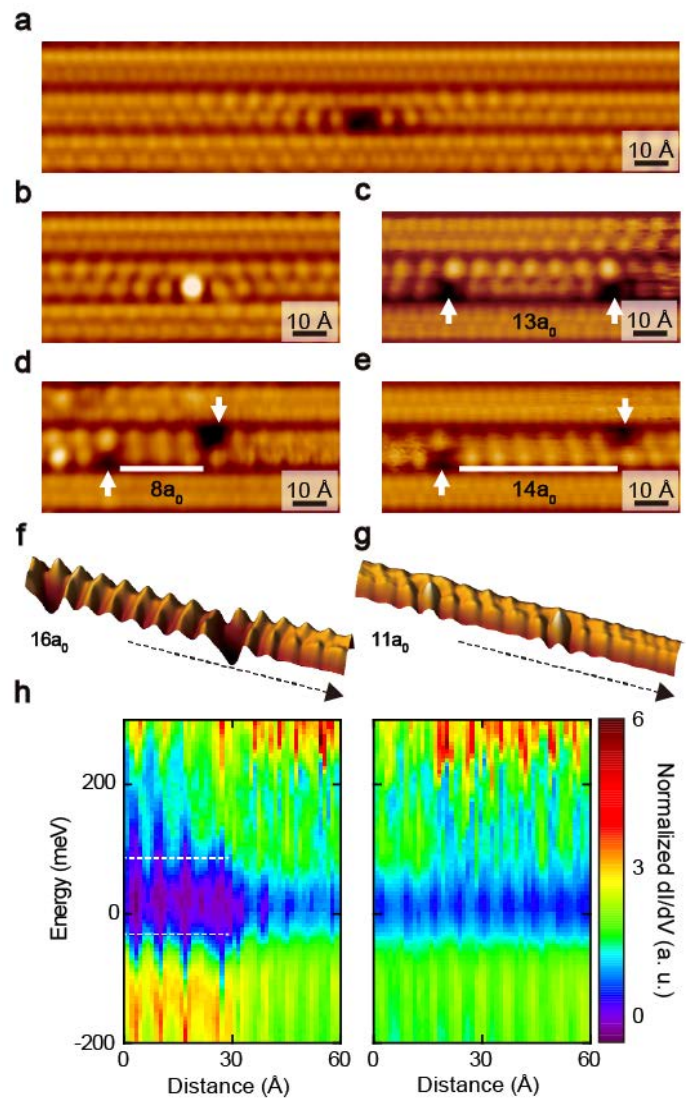

**Fig. 4**

a

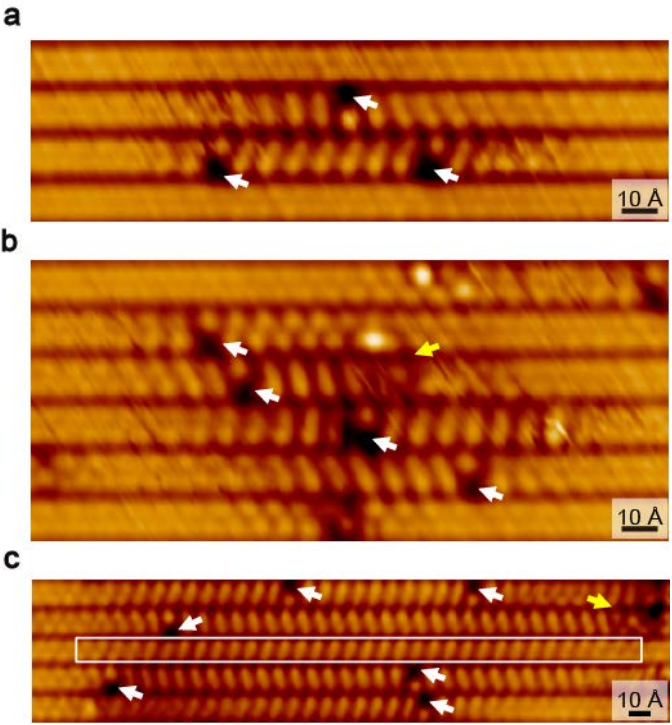

10 Å

b

10 Å

c

10 Å

18